\documentclass[printer]{tMOP2e}
\usepackage{amssymb}
\usepackage{epsfig}

\begin{document}

\title{Laser-induced nonsequential double ionization: kinematic constraints
for the recollision-excitation-tunneling mechanism}
\author{T. Shaaran and C. Figueira de Morisson Faria \\
Department of Physics and Astronomy, University College London, Gower
Street, London WC1E 6BT, United Kingdom}
\date{\today}
\maketitle

\begin{abstract}
We investigate the physical processes in which an electron, upon return to
its parent ion, promotes a second electron to an excited state, from which
it subsequently tunnels. Employing the strong-field approximation and
saddle-point methods, we perform a detailed analysis of the dynamics of the
two electrons, in terms of quantum orbits, and delimit constraints for their
momentum components parallel to the laser-field polarization. The kinetic
energy of the first electron, upon return, exhibits a cutoff slightly lower
than $10U_p$, where $U_p$ is the ponderomotive energy, as in rescattered
above-threshold ionization (ATI). The second electron leaves the excited
state in a direct ATI-like process, with the maximal energy of $2U_p$. We
also compute electron-momentum distributions, whose maxima agree with our
estimates and with other methods.
\end{abstract}

\section{Introduction}

Electron-electron correlation in strong laser fields has attracted a great
deal of attention for over a decade, in particular in the context of
laser-induced nonsequential double and multiple ionization \cite{NSDIreview}%
. For these specific phenomena, the electron-electron interaction plays a
huge role. A concrete example are the peaks in the electron momentum
distributions in nonsequential double ionization (NSDI), as functions of the
electron components $p_{n\parallel }(n=1,2)$ parallel to the laser-field
polarization \cite{Ffm2000}. Such peaks occur at nonvanishing parallel
momenta and cannot be explained by a sequential mechanism. Whilst it is
agreed upon that NSDI owes its existence to the inelastic recollision of an
electron with its parent ion, there exist several open questions related to
this recollision, such as the combined effect of the residual ionic
potential, the electron-electron interaction and the strong laser fields on
the electron-momentum distributions \cite%
{FSLB2004,Beijing2008,Eremina2004,Misha2008,sachaeckh2008}.

For instance, it can happen that the first electron, upon return, provides
the second electron with enough energy so that it is able to overcome the
binding energy of the singly ionized ion and reach the continuum. In this
case, the second electron is released by electron-impact ionization. Both
electrons leave simultaneously and lead to distributions peaked at
nonvanishing momenta, in the first and third quadrant of the plane $%
p_{1\parallel }p_{2\parallel }$ spanned by the parallel momentum components.
This particular rescattering mechanism has been extensively investigated in
the past few years, possibly for the following reasons. First, it explained
the dramatic features observed experimentally in the electron momentum
distributions, namely the peaks near the non-vanishing momenta $%
p_{1\parallel }=$ $p_{2\parallel }=\pm 2\sqrt{U_{p}}$, where $U_{p}$ is the
ponderomotive energy, and the V-shaped structure observed in the electron
momentum distributions, which is a signature of the long-range character of
the electron-electron interaction \cite%
{FSLB2004,Beijing2008,Misha2008,sachaeckh2008}. Second, especially in the
context of semi-analytic methods such as the strong-field approximation,
electron-impact ionization is easier to model than the other rescattering
mechanisms.

In the past few years, however, there has been increasing interest in
below-threshold intensities, for which the kinetic energy of the recolliding
electron is not sufficient to release a second electron by electron-impact
ionization \cite{Eremina2004,Misha2008}. In the below-threshold regime, the
second electron is promoted to an excited bound state, from which it
subsequently tunnels. This mechanism is known as the
recollision-excitation-tunneling ionization (RESI). Thereby, there is a time
delay between the ionization of the first and second electrons. While the
first electron will rescatter near a crossing of the driving field, the
second electron is expected to leave a quarter of a cycle later, i.e., near
a field maximum. Since in this case both electrons are expected to reach the
detector with opposite momenta, one anticipates that the second and fourth
quadrant of the plane $p_{1\parallel}p_{2\parallel }$ will be populated.
Apart from the below-threshold scenario, RESI is also present to a great
extent in species such as argon and helium \cite{ArvsNe,routes}.
Furthermore, recent experiments and computations involving aligned molecules
suggest that RESI plays an important role in this case \cite%
{NSDIalign,sachaeck,chineseguys,baier}.

Up to the present moment, however, there have been relatively few studies of
this specific physical mechanism, mainly in the context of classical or
semiclassical methods \cite{sachaeck,chineseguys,baier}. Indeed, in many
studies of NSDI below the threshold, electron-impact ionization has been
considered instead, either in the framework of the strong-field Coulomb
eikonal approximation \cite{Misha2008}, or in the context of a classical
model with a modified second ionization threshold \cite{Eremina2004}.

In this paper, we model this physical mechanism within the strong-field
approximation (SFA). In this specific framework, the pertaining transition
amplitude is written in terms of a semiclassical action and slowly varying
prefactors. Employing saddle-point methods, it is possible to relate the
solutions of the saddle-point equations to the classical trajectories of an
electron rescattering with its parent ion and, yet, retain quantum
mechanical features such as interference or excitation. This work is
organized as follows: In Sec. \ref{transampl}, we will give the SFA
transition amplitude for the RESI mechanism, which will be solved by
saddle-point methods. The saddle-point equations obtained will then be
analyzed in detail, and, subsequently (Sec. \ref{classicalregion}), we will
provide momentum constraints for the first and second electron. In Sec. \ref%
{results}, these constraints will then be tested against the electron
momentum distributions. Finally, in Sec. \ref{conclusions} we will conclude
the paper with a few summarizing remarks.

\section{Transition amplitude}

\label{transampl}

\subsection{General expressions}

The transition amplitude describing the recollision-excitation-tunneling
ionization (RESI) physical mechanism reads%
\begin{eqnarray}
M &=&\int_{-\infty }^{\infty }dt\int_{-\infty }^{t}dt^{\prime \prime
}\int_{-\infty }^{t^{\prime }}dt^{\prime \prime }  \label{RESI1} \\
&<&\mathbf{p}_{1}(t),\mathbf{p}_{2}(t)|\tilde{V}_{\mathrm{ion}}\tilde{U}%
(t,t^{\prime \prime })V_{12}U(t^{\prime \prime },t^{\prime })\tilde{V}|\psi
_{g}^{(1)}(t^{\prime }),\psi _{g}^{(2)}(t^{\prime })>,  \nonumber
\end{eqnarray}%
where $U(t^{\prime \prime },t^{\prime })$ and $\tilde{U}(t^{\prime \prime
},t^{\prime })$ denote the time evolution operator of the two-electron
system, $|\psi _{g}^{(1)}(t^{\prime }),\psi _{g}^{(2)}(t^{\prime })>$ is the
two-electron initial state, and $\left\vert \mathbf{p}_{1}(t),\mathbf{p}%
_{2}(t)\right\rangle $ the final two-electron continuum state. \ The
interactions $\tilde{V}=P_{cg}VP_{gg},V_{12},$ and $\tilde{V}_{\mathrm{ion}}$
=$P_{cc}V_{\mathrm{ion}}P_{ce}$ correspond to the atomic binding potential,
the electron-electron interaction and the binding potential of the singly
ionized core, respectively. \ We assume that the system is initially in a
product state of one-electron ground states, i..e., $|\psi
_{g}^{(1)}(t^{\prime }),\psi _{g}^{(2)}(t^{\prime })>=|\psi
_{g}^{(1)}(t^{\prime })>\otimes |\psi _{g}^{(2)}(t^{\prime })>$, with $|\psi
_{g}^{(n)}(t^{\prime })>=\exp [iE_{ng}t^{\prime }]|\varphi _{g}^{(2)}>.$ We
consider the length gauge and atomic units throughout.

The operators $P_{\mu \nu }$ are projectors onto the bound or continuum
subspaces. Specifically,
\begin{equation}
P_{gg}=\left\vert \varphi _{g}^{(1)},\varphi _{g}^{(2)}\right\rangle
\left\langle \varphi _{g}^{(1)},\varphi _{g}^{(2)}\right\vert  \label{P1}
\end{equation}%
is the projector onto the two-electron field-free ground state,
\begin{equation}
P_{cg}=\left\vert \mathbf{k},\varphi _{g}^{(2)}\right\rangle \left\langle
\mathbf{k},\varphi _{g}^{(2)}\right\vert  \label{P2}
\end{equation}%
projects the first electron onto the continuum state $\left\vert \mathbf{k}%
\right\rangle ,$ and keeps the second electron in the ground state $%
\left\vert \varphi _{g}^{(2)}\right\rangle $,
\begin{equation}
P_{ce}=\left\vert \mathbf{k},\varphi _{e}^{(2)}\right\rangle \left\langle
\mathbf{k},\varphi _{e}^{(2)}\right\vert  \label{P3}
\end{equation}%
projects the first electron onto the continuum state $\left\vert \mathbf{k}%
\right\rangle ,$ the second electron onto the excited state $\left\vert
\varphi _{e}^{(2)}\right\rangle $, and
\begin{equation}
P_{cc}=\left\vert \mathbf{k}_{1},\mathbf{k}_{2}\right\rangle \left\langle
\mathbf{k}_{1},\mathbf{k}_{2}\right\vert .  \label{P4}
\end{equation}%
They guarantee that the continuum and bound states remain orthogonal. For
the exact time evolution operators, this property holds. The bound-continuum
orthogonality is lost, however, if the continuum states are approximated by
Volkov states. This is one of the key assumptions within the strong-field
approximation. For details see, e.g., \cite{anatomy1}.

The time-evolution operator of the system from the tunneling time $t^{\prime
}$ of the first electron to the recollision time $t^{\prime \prime }$ was
approximated by $U(t^{\prime \prime },t^{\prime })=U_{V}^{(1)}(t^{\prime
\prime },t^{\prime })\otimes U_{g}^{(2)}(t^{\prime \prime },t^{\prime })$,
where $U_{V}^{(1)}$ is the Gordon-Volkov time-evolution operator for the
first electron and $U_{g}^{(2)}$ is the field-free time evolution operator
for the second electron in the ground state. Subsequently to the
recollision, the time evolution operator of the system was taken to be $%
\tilde{U}(t,t^{\prime \prime })=U_{V}^{(1)}(t,t^{\prime \prime })\otimes
U_{e}^{(2)}(t,t^{\prime \prime })$, where $U_{V}^{(1)}$ is the Gordon-Volkov
time-evolution operator for the first electron and $U_{e}^{(2)}$ is the
field-free time evolution operator for the second electron in the excited
state of the singly ionized ion. In particular the latter assumptions, which
are the neglect of the residual binding potential when the electron is in
the continuum and of the laser field when the electron is bound characterize
the strong-field approximation, or Keldysh-Faisal-Reiss theory.

By employing closure relations and the explicit expressions for the
Gordon-Volkov time-evolution operators, Eq. (\ref{RESI1}) can be written as
\begin{equation}
M=\int_{-\infty }^{\infty }dt\int_{-\infty }^{t}dt^{\prime \prime
}\int_{-\infty }^{t^{\prime }}dt^{\prime \prime }\int d^{3}kV_{\mathbf{p}%
_{2}e}V_{\mathbf{p}_{1}e,\mathbf{k}g}V_{\mathbf{k}g}\exp [iS(\mathbf{p}_{n},%
\mathbf{k},t,t^{\prime },t^{\prime \prime })]  \label{RESIpspace}
\end{equation}%
with the action%
\begin{eqnarray}
S(\mathbf{p}_{n},\mathbf{k},t,t^{\prime },t^{\prime \prime })
&=&-\int_{t}^{\infty }d\tau \frac{\lbrack \mathbf{p}_{2}+\mathbf{A}(\tau
)]^{2}}{2}-\int_{t^{\prime \prime }}^{\infty }d\tau \frac{\lbrack \mathbf{p}%
_{1}+\mathbf{A}(\tau )]^{2}}{2}  \nonumber \\
&&-\int_{t^{\prime }}^{t^{\prime \prime }}d\tau \frac{\lbrack \mathbf{k}+%
\mathbf{A}(\tau )]^{2}}{2}+E_{2e}(t-t^{\prime \prime })  \nonumber \\
&&+E_{2g}t^{\prime \prime }+E_{1g}t^{\prime }.
\end{eqnarray}%
Thereby, $\mathbf{A}(\tau )$ is the vector potential, the energy $E_{1g}$
denotes the first ionization potential, $E_{2g}$ the ground-state energy of
the singly ionized atom and $E_{2e}$ the energy of the state to which the
second electron is excited. The intermediate momentum of the first electron
is given by $\mathbf{k}$ and the final momenta of both electrons by $\mathbf{%
p}_{n}(n=1,2).$ Eq. (\ref{RESIpspace}) describes a physical process in which
the first electron leaves the atom at a time $t^{\prime },$ propagates in
the continuum with momentum $\mathbf{k}$ from $t^{\prime }$ to $t^{\prime
\prime },$ and upon return, gives part of the kinetic energy to the core so
that a second electron is promoted from a state with energy $E_{2g}$ to an
excited state with energy $E_{2e}.$ This electron then reaches the detector
with momentum $\mathbf{p}_{1}$. At a subsequent time $t,$ the second
electron tunnels from the excited state, reaching the detector with momentum
$\mathbf{p}_{2}$.

Within our framework, all influence of the electron-electron interaction and
of the binding potential is contained in the prefactors $V_{\mathbf{p}%
_{2}e},V_{\mathbf{p}_{1}e,\mathbf{k}g}$ and $V_{\mathbf{k}g}$. Explicitly,
they read
\begin{equation}
V_{\mathbf{p}_{2}e}=<\mathbf{p}_{2}(t)|V_{\mathrm{ion}}|\varphi _{e}^{(2)}>=%
\frac{1}{(2\pi )^{3/2}}\int d^{3}r_{2}V_{\mathrm{ion}}(\mathbf{r}_{2})e^{-i%
\mathbf{p}_{2}(t)\cdot \mathbf{r}_{2}}\varphi _{e}^{(2)}(\mathbf{r}_{2}),
\label{Vp2e}
\end{equation}%
\begin{eqnarray}
V_{\mathbf{p}_{1}e,\mathbf{k}g} &=&\left\langle \mathbf{p}_{1}(t^{\prime
\prime }),\varphi _{e}^{(2)}\right\vert V_{12}\left\vert \mathbf{k}%
(t^{\prime \prime }),\varphi _{g}^{(2)}\right\rangle =  \label{Vp1ekg} \\
&&\frac{1}{(2\pi )^{3}}\int \int d^{3}r_{2}d^{3}r_{1}e^{-i(\mathbf{p}_{1}-%
\mathbf{k})\cdot \mathbf{r}_{1}}\left[ \varphi _{e}^{(2)}(\mathbf{r}_{2})%
\right] ^{\ast }\varphi _{g}^{(2)}(\mathbf{r}_{2})V_{12}(\mathbf{r}_{1},%
\mathbf{r}_{2})  \nonumber
\end{eqnarray}%
and
\begin{equation}
V_{\mathbf{k}g}=<\mathbf{k}(t^{\prime })|V|\varphi _{g}^{(1)}>=\frac{1}{%
(2\pi )^{3/2}}\int d^{3}r_{1}e^{-i\mathbf{k(}t^{\prime })\cdot \mathbf{r}%
_{1}}V(\mathbf{r}_{1})\varphi _{g}^{(1)}(\mathbf{r}_{1}),  \label{Vk0}
\end{equation}%
where $\mathbf{k}(\tau )=\mathbf{k}+\mathbf{A}(\tau )$ and $\mathbf{p}%
_{n}(\tau )=\mathbf{p}_{n}+\mathbf{A}(\tau ),$ $\tau =t,t^{\prime
},t^{\prime \prime }.$ In the above-stated equations, $\varphi _{e}^{(2)}(%
\mathbf{r}_{2}),\varphi _{g}^{(2)}(\mathbf{r}_{2}),$ and $\varphi _{g}^{(1)}(%
\mathbf{r}_{1})$ denote the initial position-space wave functions of the
second electron in the excited state, of the second electron in the ground
state and of the first electron in the ground state, respectively. The
potentials $V(\mathbf{r}_{1})$ and $V_{\mathrm{ion}}(\mathbf{r}_{2})$
correspond to the atomic binding potential as seen by the first and second
electron, respectively. One should note that the form factor $V_{\mathbf{p}%
_{2}e}$ is formally identical that obtained for direct above-threshold
ionization, in which an electron, initially bound, reaches the detector
without rescattering \cite{BeckLew97}.

Under the additional assumption that the electron-electron interaction
depends only on the difference between the two electron coordinates, i.e., $%
V_{12}(\mathbf{r}_{1},\mathbf{r}_{2})=V_{12}(\mathbf{r}_{1}-\mathbf{r}_{2}),$
Eq. (\ref{Vp1ekg}) may be written as%
\begin{equation}
V_{\mathbf{p}_{1}e,\mathbf{k}g}=\frac{V_{12}(\mathbf{p}_{1}\mathbf{-k})}{%
(2\pi )^{3}}\int d^{3}r_{2}e^{-i(\mathbf{p}_{1}-\mathbf{k})\cdot \mathbf{r}%
_{2}}\left[ \varphi _{e}^{(2)}(\mathbf{r}_{2})\right] ^{\ast }\varphi
_{g}^{(2)}(\mathbf{r}_{2}),  \label{scattering1st}
\end{equation}%
with
\begin{equation}
V_{12}(\mathbf{k}(t^{\prime \prime }))=\int d^{3}rV_{12}(\mathbf{r})\exp [-i(%
\mathbf{p}_{1}\mathbf{-k)}\cdot \mathbf{r}]
\end{equation}%
and $\mathbf{r=r}_{1}-\mathbf{r}_{2}$. One should note that the prefactor (%
\ref{scattering1st}), resembles that obtained for high-order above-threshold
ionization, in which an electron reaches the detector after suffering one
act of rescattering \cite{BeckLew97}.

\subsection{Saddle-point analysis}

\label{saddlepoint}

In this work, we solve the transition amplitude (\ref{RESIpspace}) employing
saddle point methods. For that purpose, one must obtain the saddle-point
equations, which give the values of the variables $\mathbf{k},t,t^{\prime
\prime },t^{\prime }$ for which the action is stationary. Explicitly, these
equations are obtained from the conditions $\partial S(\mathbf{k}%
,t,t^{\prime \prime },t^{\prime })/\partial t^{\prime }=0,\partial S(\mathbf{%
k},t,t^{\prime \prime },t^{\prime })/\partial t^{\prime \prime }=0,\partial
S(\mathbf{k},t,t^{\prime \prime },t^{\prime })/\partial t=0$ and $\partial S(%
\mathbf{k},t,t^{\prime \prime },t^{\prime })/\partial \mathbf{k}=\mathbf{0.}$
This gives

\begin{equation}
\lbrack \mathbf{k}+\mathbf{A}(t^{\prime })]^{2}=-2E_{1g},  \label{saddle1}
\end{equation}

\begin{equation}
\mathbf{k}=-\frac{1}{t^{\prime \prime }-t^{\prime }}\int_{t^{\prime
}}^{t^{\prime \prime }}\mathbf{A}(\tau )d\tau ,  \label{saddle2}
\end{equation}

\begin{equation}
\lbrack \mathbf{p}_{1}+\mathbf{A}(t^{\prime \prime })]^{2}=[\mathbf{k}+%
\mathbf{A}(t^{\prime \prime })]^{2}-2(E_{2g}-E_{2e})  \label{saddle3}
\end{equation}%
and

\begin{equation}
\lbrack \mathbf{p}_{2}+\mathbf{A}(t)]^{2}=-2E_{2e}.  \label{saddle4}
\end{equation}

The saddle-point equation (\ref{saddle1}) gives the conservation of energy
at the instant $t^{\prime }$. Physically, it corresponds to tunneling
ionization of the first electron. Eq. (\ref{saddle2}) constrains the
intermediate momentum $\mathbf{k}$ of this electron so that it can return to
its parent ion. Eq. (\ref{saddle3}) expresses the fact that the first
electron returns at a time $t^{\prime \prime }$ and gives part of its
kinetic energy $E_{\mathrm{ret}}(t^{\prime \prime })=[\mathbf{k}+\mathbf{A}%
(t^{\prime \prime })]^{2}/2$ to the core, which is excited from a state with
energy $E_{2g}$ to a state with energy $E_{2e}$. This electron then reaches
the detector with final momentum $\mathbf{p}_{1}.$ Finally, a second
electron tunnels from the excited state at a subsequent time $t$, and
reaches the detector with final momentum $\mathbf{p}_{2}.$ The conservation
of energy at this instant is given by the saddle-point equation (\ref%
{saddle4}). One should note that the saddle-point Eqs. (\ref{saddle1}) and (%
\ref{saddle4}) have no real solution. In both cases, Im$[t^{\prime }]$ and Im%
$[t]$ give a rough idea of the width of the barrier and of the ionization
probability for the first and the second electron, respectively. The larger
this quantity is, the wider the barrier through which they must tunnel.

In order to compute the transition amplitude, we follow the procedure
discussed in \cite{FSB2002}, and employ the saddle point Eq. (\ref{saddle2})
to reduce the number of independent variables in the uniform approximation
used. The main difference from \cite{FSB2002} is that, in this paper, we
deal with three independent variables, i.e., $t,t^{\prime }$ and $t^{\prime
\prime },$ instead of only $t$ and $t^{\prime }.$ Apart from that, the
saddle-point equation (\ref{saddle4}) is decoupled from the remaining
saddle-point equations, so that the the ionization time $\ t$ for the second
electron can be determined independently. Physically, however, we should
guarantee that $t>t^{\prime \prime }>t^{\prime }.$

Unless stated otherwise, we consider momentum distributions for which the
perpendicular momentum components $\mathbf{p}_{n\perp }(n=1,2)$ are
integrated over. In the following, we will discuss which momentum regions
such distributions will occupy, and the physical reasons behind it.

\section{Constraints in momentum space}

\label{classicalregion}

From the saddle-point equations in the previous section, one may determine
constraints for the parallel momentum components $p_{n||}(n=1,2)$ in the
plane $p_{1\parallel }p_{2\parallel }$. These constraints will be discussed
here, and will serve as a tool to sketch an approximate shape for the
electron-momentum distributions. For simplicity, we will consider a
monochromatic field of frequency $\omega ,$ i.e., $E(t)=-dA(t)/dt=2\omega
\sqrt{U_{p}}\sin \omega t.$

Eq. (\ref{saddle4}), which corresponds to the tunneling of the second
electron, is formally identical to the saddle-point equation describing the
low-energy electrons in above-threshold ionization (ATI), the so-called
``direct electrons". In this case, an electron tunnels from a bound state
and reaches the detector without rescattering with its parent ion.

Physically, this is exactly the situation encountered for the second
electron, and will have two main consequences. Firstly, the solutions of the
saddle-point equations will be identical to those for the direct ATI
electrons \cite{RichardPhD}. For vanishing electron drift momenta, these
solutions are displaced by half a cycle, and are located at a maximum of the
field. As the momentum increases, the solutions approach each other and move
away from the maximum. Secondly, the maximal kinetic energy for the direct
ATI electrons is $2U_{p}.$ Hence, if the perpendicular components vanish, we
will have an upper and lower bound for $p_{2\parallel }.$ Explicitly, $-2%
\sqrt{U_{p}}\leq p_{2\parallel }\leq 2\sqrt{U_{p}}.$ One should note that
this in contrast to the situation discussed in our previous papers \cite%
{FSLB2004,FSLY2008}, in which the second electron is dislodged by
electron-impact ionization. In this latter case, $\pm 2\sqrt{U_{p}}$ is the
\emph{most probable} momentum $p_{2\parallel }$ with which the second
electron may leave, whereas, in the present scenario, this is the \emph{%
maximum} value for this quantity. For nonvanishing transverse momenta, this
region will remain the same. We expect, however, that there will be a large
drop in the yield. This is due to the fact that there will be an effective
increase in the potential barrier through which the electron tunnels. This
can be readily verified by writing the saddle-point equation (\ref{saddle4})
as%
\begin{equation}
\lbrack p_{2\parallel }+A(t)]^{2}=-2\tilde{E}_{2e},
\end{equation}%
with $\tilde{E}_{2e}=E_{2e}+\mathbf{p}_{2\perp }^{2}/2$.

\begin{figure}[tbp]
\begin{center}
\includegraphics[width=8.5cm,clip=true]{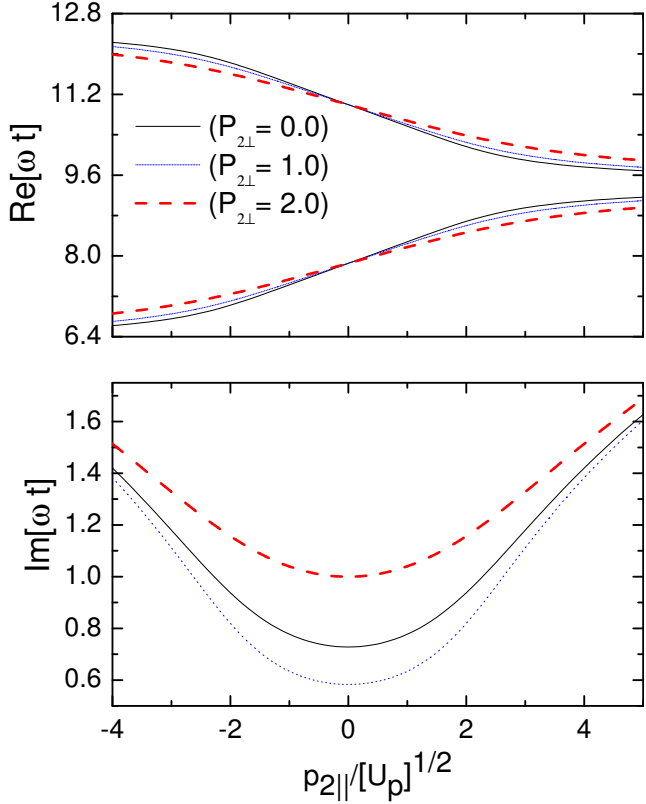} \noindent
\end{center}
\caption{Tunneling time $t$ for the second electron, as functions of its
parallel momentum $p_{2||},$ for a monochromatic field of intensity $I=1.5
\times 10^{14} \mathrm{W/cm}^{2}$ and frequency $\protect\omega =0.057$ a.u,
for several transverse momenta $p_{2\perp }$. The upper and lower panel give
the real and imaginary parts of such times, respectively. We consider that
the second electron is promoted to the $2p$ state of the Helium atom, from
which it subsequently tunnels, i.e., $E_{2e}=0.25$ a.u. and $E_{2g}=1$ a.u.}
\label{tunnelsecond}
\end{figure}
In Fig. \ref{tunnelsecond}, we plot the real and imaginary parts of such
times, as functions of the electron momentum $p_{2\parallel },$ for several
transverse momenta (upper and lower panel, respectively). In all cases, the
imaginary parts of each time $t$ in a pair are identical and exhibit a
minimum at the peak-field times $\omega t=\pi /2.$ This is expected, as
i)the two orbits behave symmetrically with respect to the laser field, and
ii)the effective potential barrier through which the electron tunnels is
narrowest for these times. As the transverse momentum $\mathbf{p}_{2\perp}$
becomes larger in absolute value, we see an increase in Im$[t]$. This is
consistent with the fact that the potential barrier widens in this case.

Eq. (\ref{saddle3}), on the other hand, has a similar form as the
saddle-point equation describing the rescattered electrons in ATI \cite%
{FSB2002}, apart from the energy difference $E_{2g}-E_{2e}$ on the
right-hand side. Physically, this is somehow expected, as in both cases the
first electron leaves immediately after rescattering. The difference is
that, while in ATI the rescattering is elastic, in NSDI part of the
electron's kinetic energy is used to excite the core. For ATI, the maximal
energy with which the electron rescatters is $10U_{p},$ and corresponds to
backscattered electrons. Therefore, to first approximation, this will be
employed to compute the upper bound for the parallel electron momentum $%
p_{1\parallel }$. Clearly, this kinetic energy will be subtracted by $%
E_{2g}-E_{2e}$ and is slightly smaller.

Apart from that, there is also a minimal energy for the first electron to
excite its parent ion and still rescatter. This is given by the condition $%
E_{2g}-E_{2e}=[\mathbf{k}+\mathbf{A}(t^{\prime \prime })]^{2}/2$, and
implies a vanishing right-hand side on Eq. (\ref{saddle3}). For vanishing
perpendicular momentum, this will lead to $p_{1\parallel }=-A(t^{\prime
\prime })$. Since, to first approximation, the electron returns at a field
crossing, this implies that $-2\sqrt{U_{p}}\lesssim p_{1\parallel }\lesssim 4%
\sqrt{U_{p}}.$ For the orbits leading to the mirror image of the
distribution with respect to the reflection $(p_{1\parallel },$ $%
p_{2\parallel })\rightarrow $ $(-p_{1\parallel },-p_{2\parallel })$, the
constraint upon the parallel momentum of the first electron will be $-4\sqrt{%
U_{p}}\lesssim p_{1\parallel }\lesssim 2\sqrt{U_{p}}.$ For these latter
orbits, the times $t^{\prime },t^{\prime \prime }$ and $t$ are displaced by
half a cycle. A nonvanishing transverse momentum component $\mathbf{p}%
_{1\perp }$ will lead to lower maximal and minimal momenta.

\begin{figure}[tbp]
\begin{center}
\includegraphics[width=13cm,clip=true]{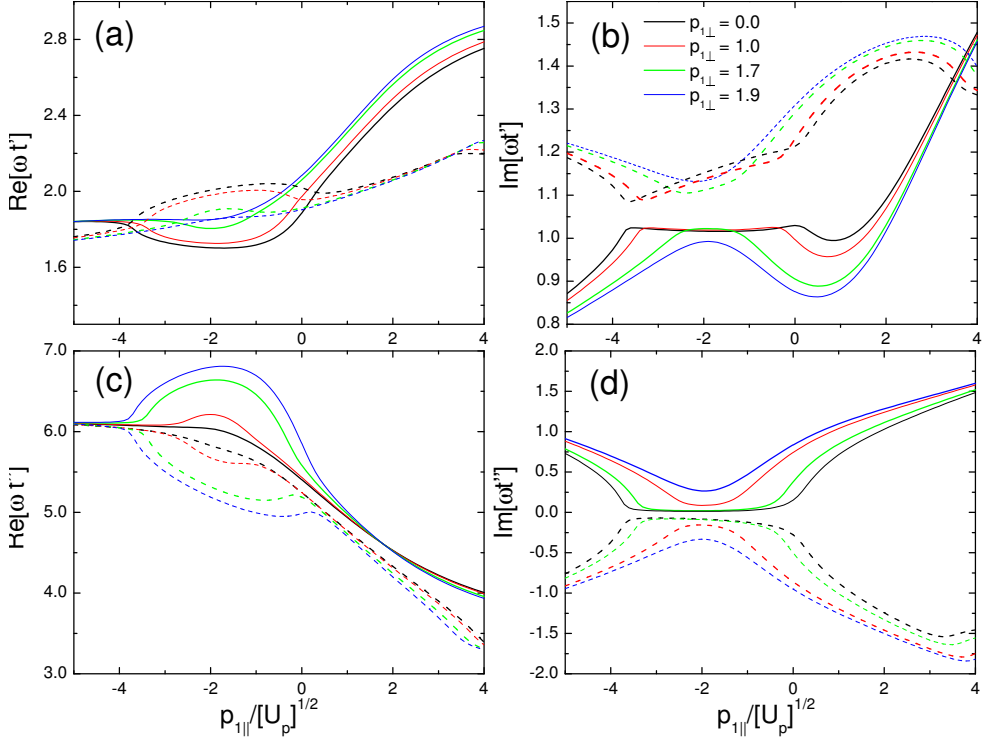} \noindent
\end{center}
\caption{Tunneling and rescattering times for the first electron, as
functions of its parallel momentum $p_{1||}$. Panels (a) and (b) give the
real and imaginary parts of the tunneling time $t^{\prime }$, respectively,
and panels (c) and (d) depict the the real and imaginary parts of the
rescattering time time $t^{\prime \prime }$. We consider that the electron
tunnels from the $2s$ state of the Helium atom, i.e., $E_{1g}=0.92$ a.u.,
and rescatters with the $1s$ state of $He^{+},$ i.e., $E_{2g}=1$ a.u.
Thereby the returning electron gives part of its kinetic energy to excite a
second electron to the state $E_{2e}=0.25$ a.u. The dashed and solid lines
correspond to the short and long orbits, respectively.}
\label{timesfirst}
\end{figure}
In Fig. \ref{timesfirst}, we display the real and imaginary part of the
ionization [panels (a) and (b), respectively] and rescattering times [panels
(c) and (d), respectively] for the first electron. We consider the shortest
orbits for the returning electron. The remaining sets of orbits are strongly
suppressed due to wave-packet spreading. By associating the real parts of $\
t^{\prime }$ and $t^{\prime \prime }$ with the classical trajectories of an
electron in a laser field, one may identify a longer and a shorter orbit,
along which the first electron returns. These orbits practically coalesce
for two specific values of $p_{1\parallel },$ namely the minimum and the
maximum momenta for which the rescattering process described by the
saddle-point equation (\ref{saddle3}) has a classical counterpart. Beyond
these momenta, the yield decays exponentially. For vanishing transverse
momentum $p_{1\perp },$ these cutoffs are near $-4\sqrt{U_{p}}$ and $2\sqrt{%
U_{p}}$, as predicted by our estimates. As $p_{1\perp }$ increases, the
classically allowed region shrinks and gets very localized near $%
p_{1\parallel }=-2\sqrt{U_{p}}.$ For the parameters considered here, this
corresponds to the situation in which the electron returns at a crossing of
the field. Finally, for very large transverse momenta, this region
disappears.

The imaginary parts of the times $\ t^{\prime }$ and $t^{\prime \prime }$,
displayed in Figs. \ref{timesfirst}.(b) and \ref{timesfirst}.(d), confirm
this physical interpretation. In fact, they show that, for the rescattering
times, Im[$t^{\prime \prime }$] essentially vanishes between the momenta for
which the real parts Re[$t^{\prime \prime }$] coalesce. Physically, this
means that, in this region, rescattering is classically allowed. Beyond this
region, Im[$t^{\prime \prime }$] increases abruptly, which indicates that
the classically forbidden region has been reached. \ In this context, it is
worth mentioning that, even if there is no classically allowed region, Im[$%
t^{\prime \prime }$] exhibits a minimum near $p_{1\parallel }=-2\sqrt{U_{p}}$%
. This is due to the fact that rescattering is most probable for this
specific momentum. A similar behavior has been observed in \cite{FB2003} for
electron-impact ionization.

\begin{figure}[tbp]
\begin{center}
\noindent \includegraphics[width=9cm]{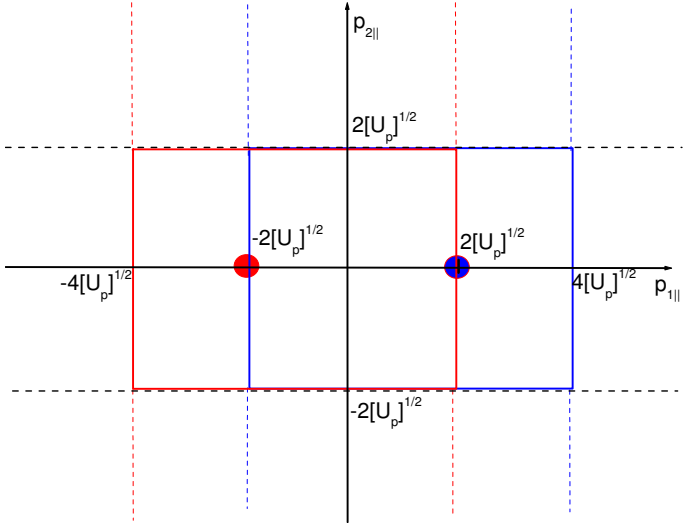}
\end{center}
\caption{{}Schematic representation of the regions of the parallel momentum
plane populated by the recollision-excitation-tunneling ionization
mechanism, highlighted as the rectangles in the figure. The dashed black
lines indicate the constraints in the parallel momentum $p_{2||}$ of the
second electron, while the remaining dashed lines depict the limits for the
parallel momentum $p_{1||}$ of the first electron. The circles indicate the
expected maxima of the electron momentum distributions. The blue and red
colors indicate different sets of trajectories, whose start and recollision
times are separated by half a cycle of the field. In our estimates, we
considered vanishing transverse momenta, so that the constraints provided
constitute an upper bound for this region. The above plot has not been
symmetrized with respect to the indistinguishability of the two electrons.}
\label{constraintsp1p2}
\end{figure}
The imaginary part Im[$t^{\prime }$] of the start time of the first
electron, on the other hand, is always non-vanishing. This is not
surprising, as tunneling has no classical counterpart. They are, however,
approximately constant between the lower and upper cutoff momenta.

By solving the saddle-point Eqs. (\ref{saddle3}) and (\ref{saddle4}), we
verified that, if we consider the physically relevant parameters, namely
that $t$ $\sim t^{\prime \prime }+T/4,$ where $T=2\pi /\omega $ \ denotes a
laser-field cycle, then $p_{2||}$ will be predominantly positive and $%
p_{1||} $ will be predominantly negative. Conversely, for the orbits whose
start, rescattering and tunneling times are displaced by half a cycle, $%
p_{2||}$ will be mainly negative and $p_{1||}$ will be predominantly
positive. Therefore, one expects that the distributions will be mainly
concentrated in the second and fourth quadrants of the parallel momentum
plane.

In Fig. \ref{constraintsp1p2}, we summarize the information
discussed above, and provide a schematic representation of the
momentum regions occupied in the RESI process. In particular, we
expect the distributions to exhibit maxima near the points
$(p_{1||},p_{2||})=(\pm 2\sqrt{U_{p}},0).$ In a real-life situation,
since both electrons are indistinguishable, one would expect maxima
also at $(p_{1||},p_{2||})=(0,\pm 2\sqrt{U_{p}})$.

\section{Electron momentum distributions}

\label{results}

In this section, we compute electron-momentum distributions employing Eq. (%
\ref{RESIpspace}), under the assumption that the prefactors $V_{\mathbf{p}%
_{2}e},V_{\mathbf{p}_{1}e,\mathbf{k}g}$ and $V_{\mathbf{k}g}$ are constant.
This removes any momentum bias that may arise from such prefactors, and
therefore provides a clearer picture of how the momentum-space constraints
affect such distributions. The transverse momentum components are integrated
over.

\begin{figure}[tbp]
\begin{center}
\noindent \includegraphics[width=13cm]{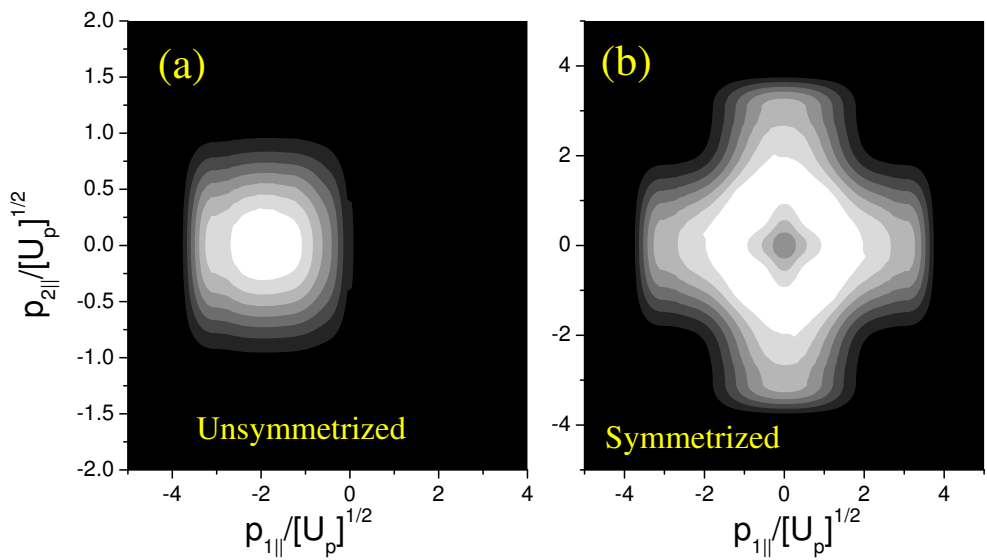}
\end{center}
\caption{Electron momentum distributions for Helium ($E_{1g}=0.92$ a.u., $%
E_{2g}=1$ a.u. and $E_{2e}=0.25$) in a linearly polarized, monochromatic
field of frequency $\protect\omega=0.057$ a.u. and intensity $I=1.5\times%
\mathrm{W/cm^2}$. Panel (a) displays only the contributions from the sets of
orbits starting at $0<t^{\prime}<T/2$, while panel (b) depicts also the
contributions from the other half-cycle of the field. In panel (b), the
distributions have also been symmetrized with respect to the exchange $%
\mathbf{p}_1\leftrightarrow \mathbf{p}_2$ }
\label{cross}
\end{figure}
Fig. \ref{cross} depicts such distributions. In panel (a), we consider only
that the first electron is released in $0<t^{\prime}<T/2$, where $%
T=2\pi/\omega$ denotes a cycle of the external driving field, while in panel
(b) we also consider the contributions from $t^{\prime}\rightarrow
t^{\prime}\pm T/2$, $t^{\prime\prime}\rightarrow t^{\prime\prime}\pm T/2$
and $t\rightarrow t\pm T/2$. Furthermore, in the latter case, we also
symmetrize the distributions with respect to $\mathbf{p}_1\leftrightarrow
\mathbf{p}_2$, as the two electrons are indistinguishable. We have
considered the parameters for Helium, corresponding to the situation in
which an electron initially in $1s$ was released and promoted a second
electron to the $2p$ state.

In Fig. \ref{cross}.(a), one clearly sees that the distributions are
brightest along the axis $p_{2||}=0$. This is expected, as the emission of
the second electron is most probable at a field maximum.  For this time, the
electron momentum vanishes. Apart from that, the  distribution is longer in
the $p_{1||}$ direction. This is  expected, as the cutoff momenta is higher
in this case. Finally,  the distributions also exhibit a maximum at $%
p_{1||}=-2\sqrt{U_p}$,  in agreement with the above-defined constraints.
Upon symmetrization  [Fig. \ref{cross}.(b)], we obtain distributions highly
concentrated  along the momentum axis $p_{1||}=0$ and $p_{2||}=0$. These
distributions also exhibit a ring-shaped maximum around the  origin of the $%
p_{1||}p_{2||}$ plane. These results show that the  momentum regions
populated by the RESI mechanism are much lower  than those populated if the
second electron is released by  electron-impact ionization, in agreement
with other results  reported in the literature \cite{ArvsNe,routes}.

\section{Conclusions}

\label{conclusions} The main conclusion to be inferred from this work is
that the recollision-excitation-ionization mechanism, which is becoming
increasingly studied due to its importance for NSDI of molecules and at
threshold intensities, can be understood as rescattered above-threshold
ionization (ATI) for the first electron, followed by direct ATI for the
second electron. The kinematic constraints imposed by both processes lead to
cross-shaped electron momentum distributions, localized at the axis $%
p_{1\parallel}=0$ or $p_{2\parallel}=0$, and centered at $%
p_{1\parallel}=p_{2\parallel}=0$.

The fact that these distributions are concentrated in the low momentum
regions is not surprising. Physically, much less energy is required to
promote an electron to an excited state, from which it subsequently tunnels,
than to provide the second electron with enough energy so that it may
overcome the second ionization potential and reach the continuum, as in
electron-impact ionization \cite{FSLB2004,Misha2008}. Furthermore, since for
the recollision-excitation-tunneling mechanism there is a time delay between
the ionization of the first and second electron, the second and fourth
quadrants of the plane spanned by the parallel momentum components $p_{n||}$
are populated \cite{sachaeck,chineseguys}. This is not the case in
electron-impact ionization, for which both electrons leave simultaneously.
In contrast to the results reported in \cite{sachaeck,chineseguys}, however,
we did not observe a localization of the distributions only in such regions.
Such an effect is possibly due to the influence of the long-range tail of
the Coulomb potential, and is presently under investigation.

\textbf{Acknowledgements} This work has been financed by the UK EPSRC
(Advanced Fellowship, Grant no. EP/D07309X/1 and DTA studentship). We thank
A. Emmanouilidou and M. Ivanov for useful discussions.

\end{document}